\documentstyle[aps,prd,psfig,floats,twocolumn]{revtex}
\preprint{DAMTP-2002-xyz}
\date{August 2002}
\tighten
\begin{document}
\title{Suppression of Entropy Perturbations in Multi--Field Inflation on the Brane}
\author{P.R. Ashcroft, C. van de Bruck and A.--C. Davis}
\address{Department of Applied Mathematics and Theoretical Physics, 
Center for Mathematical Sciences,\\
University of Cambridge, Wilberforce Road, Cambridge CB3 0WA, U.K.}
\noindent
\maketitle
\begin{abstract}
At energies higher than the brane tension, the dynamics of a 
scalar field rolling down a potential are modified relative to 
the predictions of General Relativity. The modifications imply, 
among other things, that steeper potentials can be used to drive
an epoch of slow--roll inflation. We investigate 
the evolution of entropy and adiabatic 
modes during inflation driven by two scalar fields confined on the brane. 
We show that the  
amount of entropy perturbations produced during inflation 
is suppressed compared to the predictions made by General Relativity. 
As a consequence, the initial conditions do not matter in multiple
field inflation in brane worlds if inflation is driven at energies
much higher than the brane tension.
\end{abstract}
\vspace{0.3cm}
\noindent DAMTP-2002-107
\vspace{0.3cm}

Inflation provides an explanation for the origin 
of structures in the universe. In its simplest form, inflation is
driven by a single scalar field, called the inflaton, which slowly rolls 
down a potential. Perturbations in the inflaton field are 
adiabatic and obey Gaussian statistics [\ref{liddlelyth}]. 
However, this picture might be over simplistic as there is no reason 
to believe that inflation is driven solely by a single scalar field. 
Instead, the possibility that inflation is driven by a system of
several scalar fields must be considered. In particular, perturbations
are, in general, no longer purely adiabatic. 
Entropy (or isocurvature) perturbations can be generated. 
In addition, non--Gaussian fluctuations are now possible. 
In the case of multiple field inflation, there are no unique background 
trajectories for the fields and therefore predictions of density
perturbations could depend on initial conditions [\ref{bellido}].

Current data do not exclude an additional entropy mode 
[\ref{durrer},\ref{luca}]. Future experiments such as the 
Microwave Anisotropy Probe (MAP), the Planck Surveyor, the 
2 Degree Field Galaxy Survey (2dF) and Sloan Digital Sky Survey
(SLOAN) will constrain deviations from the standard 
predictions of single scalar field inflation. These prospects 
motivated recent work to consider multi--field inflation and the 
consequences for density perturbations [\ref{gordon}--\ref{bartolo}].

In the case of brane worlds based on the model of Randall and 
Sundrum [\ref{randall}], inflation driven by a scalar field 
confined on the brane has been considered (see e.g. [\ref{wands}--\ref{lidsey}]. 
It was shown that the correction of Einstein's equation at high 
energies lead to a suppression of the usual slow roll parameters 
due to the enhanced damping of the scalar
field. This in turn implies that the spectral index of the 
spectrum of perturbations is nearer one for a given potential compared
to the predictions of General Relativity. In this letter we 
point out that the enhanced damping implies that the 
generation of entropy perturbations is suppressed in the case of 
multiple--field inflation on the brane for a given potential. 
We also deduce that steeper potentials are needed to 
create the same amount of entropy modes as in usual 
four--dimensional gravity. Furthermore, any transfer of
non--Gaussianity from the entropy modes to the adiabatic modes 
is suppressed. If inflation is driven at energies much higher than the 
brane tension, the results of density perturbations in multiple field
inflation become independent of the initial conditions. 

{\it General Relativity:} Before we consider inflation on a brane, 
we review briefly the results obtained in General Relativity
[\ref{wandsriotto}]. It turns out the form of the equations are not
modified in the case of the brane world model we consider. 

To be precise, we consider the case of two scalar fields, $\phi$ and
$\chi$, which are slowly rolling down 
some potential $V(\phi,\chi)$. The equations of motion are governed
by the Friedmann equation and Klein--Gordon equations:
\begin{eqnarray}
H^2 &=& \frac{8\pi G}{3}\rho, \label{normal friedmann} \\
\ddot \phi &+& 3H\dot \phi + V_{\phi} = 0,\label{kleingordon1} \\
\ddot \chi &+& 3H\dot \chi + V_{\chi} = 0, \label{kleingordon2}
\end{eqnarray}
where $\rho$ is the energy density of {\it both} fields. 
We assume that the fields are slow--rolling, implying that the
parameters 
\begin{eqnarray}
\epsilon_{\phi} &\equiv& \frac{1}{16\pi
G}\left(\frac{V_{\phi}}{V}\right)^2 , \vspace{0.5cm} \epsilon_{\chi} \equiv \frac{1}{16\pi
G}\left(\frac{V_{\chi}}{V}\right)^2 \nonumber \\
\eta_{\phi\phi} &\equiv& \frac{1}{8\pi
G}\left(\frac{V_{\phi\phi}}{V}\right), \vspace{0.5cm} \eta_{\phi\chi}
\equiv \frac{1}{8\pi G}\left(\frac{V_{\phi\chi}}{V}\right) \label{slowrollgr}\\
\eta_{\chi\chi} &\equiv& \frac{1}{8\pi
G}\left(\frac{V_{\chi\chi}}{V}\right) \nonumber 
\end{eqnarray}
are small.
A convenient formalism to study entropy perturbations 
was presented in [\ref{gordon}] which 
we will use here. Instead of working with the fields $\phi$ and 
$\chi$ it is useful to perform a field rotation as follows:
\begin{eqnarray}
\delta \sigma &=& (\cos \theta) \delta\phi   +   (\sin \theta) \delta\chi \\
\delta s &=& -  (\sin \theta) \delta\phi + (\cos \theta) \delta\chi,
\end{eqnarray}
with
\begin{equation}
\cos \theta = \frac{\dot\phi}{\sqrt{\dot\phi^2 + \dot\chi^2}}, \hspace{1cm}
\sin \theta = \frac{\dot\chi}{\sqrt{\dot\phi^2 + \dot\chi^2}}.
\end{equation}
$\sigma$ is called the adiabatic field and $s$ is called the 
entropy field. The meaning of the names of the fields becomes 
clear when one considers their fluctuations. 

The line--element for arbitrary scalar perturbations of the
Robertson--Walker metric for a spatially flat universe 
reads (using the notation of [\ref{gordon}]) 
\begin{eqnarray}
ds^2 &=& -(1+2A)dt^2 + 2 a^2 B_{,i} dx^i dt \nonumber \\
&+& a^2 \left[(1-2\psi)\delta_{ij} + 2 E_{,ij}\right]dx^i dx^j.
\end{eqnarray}
The gauge--invariant curvature perturbation, defined as 
\begin{equation}
{\cal R} = \psi + \frac{H \delta \rho}{\dot \rho}, \label{curvpert}
\end{equation}
is, on very large scales, constant for purely adiabatic
perturbations (see e.g. [\ref{malik}] and references therein). 
However, entropy perturbations are a source for the curvature perturbation
(\ref{curvpert}). In addition, the entropy perturbation between two 
species $A$ and $B$, defined as 
\begin{equation}
{\cal S} = \frac{\delta n_A}{n_A} - \frac{\delta n_B}{n_B},
\end{equation}
where $n_i$ are the number densities of the particle species $i$, can 
evolve in time, even on superhorizon scales. Therefore it was argued 
[\ref{wandsriotto}],
that on very large scales in general we have the following equations
describing the evolution of ${\cal R}$ and ${\cal S}$:
\begin{eqnarray}
\dot {\cal R} &=& \alpha H {\cal S} \label{eomR}\\
\dot {\cal S} &=& \beta H {\cal S} \label{eomS}
\end{eqnarray}
For the case of the two slow-rolling scalar fields and in the spatial
flat gauge ($\psi=0$), ${\cal R}$ and ${\cal S}$ are given by
\begin{equation}
{\cal R} \approx \frac{H\left( \dot\phi \delta\phi + \dot\chi
\delta\chi\right)}{\dot\phi^2 + \dot\chi^2} = \frac{H \delta
\sigma}{\dot \sigma}
\end{equation}
and 
\begin{equation}
{\cal S} = \frac{H\left( \dot\phi \delta\chi - \dot\chi
\delta\phi\right)}{\dot\phi^2 + \dot\chi^2} = \frac{H \delta s}{\dot \sigma}
\end{equation}
Fluctuations in the field $\sigma$ are adiabatic perturbations, 
whereas fluctuations in $s$ are entropic perturbations. On very large
scales ($k\ll aH$) and in the flat gauge
the evolution of fluctuations are described as [\ref{gordon}]:
\begin{eqnarray}
(\delta \sigma)^{..} + 3 H (\delta \sigma)^. + 
\left( V_{\sigma\sigma} - \dot\theta^2\right)\delta\sigma = \label{adiafield}\\ 
-2 V_\sigma A + \dot\sigma \dot A + 2(\dot\theta\delta s)^. - 
2\frac{V_\sigma}{\dot \sigma}\dot \theta \delta s \nonumber
\end{eqnarray}
and
\begin{eqnarray}
(\delta s)^{..} + 3 H (\delta s)^. + 
\left( V_{ss} - \dot\theta^2\right)\delta s  = \label{entrofield}\\
-2\frac{\dot\theta}{\dot\sigma}\left[ \dot\sigma((\delta \sigma)^. -
\dot\sigma A) -\ddot \sigma \delta\sigma\right].\nonumber
\end{eqnarray} 
The metric perturbation $A$ can be obtained from Einstein's equation 
and is given, in the flat gauge, by
\begin{equation}
H A =  4\pi G \left(\dot\phi \delta\phi + \dot\chi \delta\chi\right)
\end{equation}

It was shown in [\ref{wandsriotto}] that for the case of two 
slow--rolling scalar fields, $\alpha$ and $\beta$ in (\ref{eomR}) and 
(\ref{eomS}) are given in terms of the slow roll parameters
\begin{eqnarray}
\alpha &=& -2\eta_{\sigma s}, \label{alpha} \\
\beta &=& -2\epsilon + \eta_{\sigma\sigma} - \eta_{ss} \label{beta},
\end{eqnarray}
and are, therefore, specified by the potential $V(\phi,\chi)$. In the 
last two equations, the slow roll parameter are constructed from the 
usual slow--roll parameter for $\phi$ and $\chi$ (\ref{slowrollgr}) and are given by
\begin{equation}\label{epsilongr}
\epsilon = \frac{1}{16\pi G}\left(\frac{V_{\sigma}}{V}\right)^2 
\approx \epsilon_\phi + \epsilon_\chi. 
\end{equation}
and
\begin{eqnarray}
\eta_{\sigma\sigma} &=& \eta_{\phi\phi}\cos^2\theta + 
2\eta_{\phi\chi}\cos\theta\sin\theta + 
\eta_{\chi\chi} \sin^2 \theta, \nonumber \\
\eta_{s s} &=& \eta_{\phi\phi}\sin^2\theta -  
2\eta_{\phi\chi}\cos\theta\sin\theta + 
\eta_{\chi\chi} \cos^2 \theta,\label{etagr} \\
\eta_{\sigma s} &=&
(\eta_{\chi\chi}-\eta_{\phi\phi})\sin\theta\cos\theta 
+ \eta_{\phi\chi}(\cos^2 \theta - \sin^2 \theta). \nonumber
\end{eqnarray}
The time evolution of ${\cal R}$ and ${\cal S}$ between horizon
crossing and some later time are given by
\begin{eqnarray}
\left(\begin{array}{c}
{\cal R} \\
{\cal S} \end{array} \right) = 
\left( \begin{array}{cc} 
1 & T_{{\cal R}{\cal S}} \\
0 & T_{{\cal S}{\cal S}} \end{array} \right) \left(\begin{array}{c}
{\cal R} \\
{\cal S} \end{array} \right)_*,
\end{eqnarray}
where the asterisk marks the time of horizon crossing. The transfer
functions $T_{{\cal R}{\cal S}}$ and $T_{{\cal S}{\cal S}}$ are given
by 
\begin{eqnarray} 
T_{{\cal S}{\cal S}}(t,t_*) &=& \exp\left( \int_{t_*}^t \beta(t') 
H(t') dt' \right)  \nonumber \\
T_{{\cal R}{\cal S}}(t,t_*) &=& \int_{t_*}^t \alpha(t')T_{{\cal
S}{\cal S}}(t_*,t') H(t') dt'. \label{transfer}
\end{eqnarray}
Thus, in the case of two slow rolling scalar fields, the transfer
functions are completely specified by the potential through the 
slow--roll parameter. In the following we would like to find the 
changes of the evolution of adiabatic and entropy perturbations
in brane world theories. 

{\it Brane World in AdS:} In the case of a 
brane embedded in an Anti--de Sitter spacetime,
Einstein's equation is not valid at high energies. Here, 
modifications become important in the form of a term
which is quadratic in $T_{\mu\nu}$, the energy--momentum tensor of 
matter confined on the brane. In addition, the gravitational field 
in the bulk is encoded in the projection of the five--dimensional 
Weyl--tensor onto the brane. It was shown that the Einstein--equation 
has the from [\ref{shiro}]
\begin{equation}\label{einstein}
G_{\mu\nu} = \kappa_4^2 T_{\mu\nu} + \kappa_5^2 \Pi_{\mu\nu} - E_{\mu\nu},
\end{equation}
where $\Pi_{\mu\nu}$ is the term quadratic in $T_{\mu\nu}$ and $E_{\mu\nu}$ is 
the projected Weyl tensor. 
For the case of a homogeneous and isotropic expanding brane embedded
in an Anti--de Sitter space ($E_{\mu\nu}=0$), the Friedmann equation 
is given by
\begin{equation}
H^2 = \frac{8\pi G}{3}\rho\left[1 + \frac{\rho}{2\lambda}\right].
\end{equation}
In this equation $\lambda$ is the (intrinsic) brane tension and
$\rho$ is the total energy density of the scalar 
fields\footnote{We have assumed that the four--dimensional 
cosmological constant is negligible.}. The equations of motion for 
scalar fields confined on the brane are given by the usual Klein--Gordon
equations (\ref{kleingordon1}) and (\ref{kleingordon2}).

It was shown in [\ref{wands}] that slow roll inflation driven by a 
single scalar field demands that the two slow--roll parameters
\begin{eqnarray}
\epsilon &\equiv& \frac{1}{16\pi G}\left(\frac{V'}{V}\right)^2\left[ 
\frac{4\lambda(\lambda + V)}{(2\lambda + V)^2} \right] \label{epsilon}\\
\eta &\equiv& \frac{1}{8\pi G}\left(\frac{V''}{V}\right)\left[
\frac{2\lambda}{2\lambda + V}\right]\label{eta}
\end{eqnarray}
are small. The number of e--folds is enhanced for a given potential
compared to the usual four--dimensional case. It was also shown that 
the amplitude of scalar perturbations is enhanced relative to the
standard result in four dimensions for a given potential. 

In the case of the two scalar fields $\phi$ and $\chi$, the set of slow--roll 
parameters is given by
\begin{eqnarray}
\epsilon_{\phi} &\equiv& \frac{1}{16\pi G}\left(\frac{V_{\phi}}{V}\right)^2\left[ 
\frac{4\lambda(\lambda + V)}{(2\lambda + V)^2} \right], \nonumber \\
\epsilon_{\chi} &\equiv& \frac{1}{16\pi G}\left(\frac{V_{\chi}}{V}\right)^2\left[ 
\frac{4\lambda(\lambda + V)}{(2\lambda + V)^2} \right], \nonumber \\
\eta_{\phi\phi} &\equiv& \frac{1}{8\pi G}\left(\frac{V_{\phi\phi}}{V}\right)\left[
\frac{2\lambda}{2\lambda + V}\right] ,\\
\eta_{\phi\chi} &\equiv& \frac{1}{8\pi G}\left(\frac{V_{\phi\chi}}{V}\right)\left[
\frac{2\lambda}{2\lambda + V}\right], \nonumber \\
\eta_{\chi\chi} &\equiv& \frac{1}{8\pi G}\left(\frac{V_{\chi\chi}}{V}\right)\left[
\frac{2\lambda}{2\lambda + V}\right], \nonumber
\end{eqnarray}
which implies the corresponding changes for the slow--roll parameters 
for the adiabatic field $\sigma$ and $s$. 
The changes in perturbations on brane worlds are two--fold. First, 
we perturb the spacetime away from Anti--de Sitter, meaning that 
the projected Weyl--tensor no longer vanishes. i.e. $\delta
E_{\mu\nu} \neq 0$. However, in the case of a de Sitter brane and 
on large scales, it was shown that the Weyl contribution is
negligible (e.g. [\ref{roy}]). We therefore neglect that term. The second contribution 
is quadratic in $T_{\mu\nu}$, which we do not neglect. It implies
that the metric perturbation $A$ is now given by [\ref{langlois}]
\begin{equation}
HA = 4\pi G\left(1 +
\frac{\rho}{\lambda}\right)\left(\dot\phi\delta\phi + 
\dot\chi\delta\chi \right). 
\end{equation}
Using (\ref{adiafield}) and (\ref{entrofield}), it is not difficult to 
show that these changes imply that
\begin{eqnarray}
H^{-1} (\delta \sigma)^. \simeq \left( 2\epsilon
-\eta_{\sigma\sigma} \right)\delta \sigma - 2\eta_{\sigma s}\delta s
\end{eqnarray}
and
\begin{eqnarray}
H^{-1} (\delta s)^. \simeq -\eta_{ss} \delta s,
\end{eqnarray}
on large scales in the slow roll regime.
These are exactly the same expressions found in [\ref{wandsriotto}]
for General Relativity. This implies that the evolution 
of ${\cal R}$ and ${\cal S}$ are given by the expressions 
\begin{eqnarray}
\dot{{\cal R}} &\simeq& - 2 H \eta_{\sigma s}{\cal S} \\
\dot{{\cal S}} &\simeq& \left[ -2\epsilon -\eta_{ss} 
+ \eta_{\sigma\sigma} \right]H{\cal S}
\end{eqnarray}
Thus, we have found that the equations have the same form as in usual 
cosmology based on General Relativity. Therefore, the parameters 
$\alpha$ and $\beta$ are the same as in eq. (\ref{alpha}) and
(\ref{beta}), but for a given potential they 
are reduced by a factor $\approx \lambda/V$ at high energies on the
brane. At very high energies, i.e. $V \gg \lambda$, we have that 
the ratios $\dot{{\cal R}}/{\cal S}$ and $\dot{{\cal S}}/{\cal S}$
are suppressed by a factor $\approx \sqrt{\lambda/V}$. That means that 
the transfer functions (\ref{transfer}) are significantly modified in 
that case. Indeed, for very large energies we have 
$T_{{\cal S}{\cal S}} \rightarrow 1$ and $T_{{\cal R}{\cal S}} 
\rightarrow 0$ as $V/\lambda \rightarrow \infty$, 
meaning that entropy perturbations have less influence
on adiabatic perturbations and even if entropy
perturbations are created, the final correlation between the 
adiabatic and entropy modes would be suppressed. 

Our findings imply that, even in models with step potentials (see [\ref{liddle}]) 
and multiple scalar fields, the production of entropy perturbations is
suppressed, as long as the energies are much higher than the brane
tension.

It is not difficult to understand the physical reason for the results 
obtained so far. From equation (\ref{adiafield}) and
(\ref{entrofield}) we see that the angle $\theta$ merely has to 
vary in order for the fields $s$ and $\sigma$ to interact\footnote{If
$\theta$ is time varying, the background trajectory is curved.}. 
For the case of the slow--rolling scalar fields on the brane, 
the equation of motion for the angle $\theta$ in the slow roll regime 
is given by 
\begin{equation}
\dot\theta \simeq - \eta_{\sigma s} H.
\end{equation}
This equation holds also for General Relativity (see
[\ref{wandsriotto}]). For a given potential, this equation implies 
that, for brane world inflation at high energies, $\theta$ is almost 
constant, and therefore that the background trajectory will quickly 
follow a straight line (see figure). 

For the extreme case $V/\lambda \rightarrow \infty$, the equation 
above implies that $\dot\theta \rightarrow 0$. In that case there 
is no efficient transfer of energy between $s$ and $\sigma$. Instead, 
$s$ evolves like a background field which does not influence 
perturbations in $\sigma$ and the gravitational potential. Note 
that this also means that any transfer of non--Gaussianity from 
the entropy field $s$ to the gravitational potential and/or the 
adiabatic field $\sigma$ is strongly suppressed (see [\ref{uzan}] 
for a recent discussion). 

\begin{figure}[!h]
\hspace{.5cm}\psfig{file=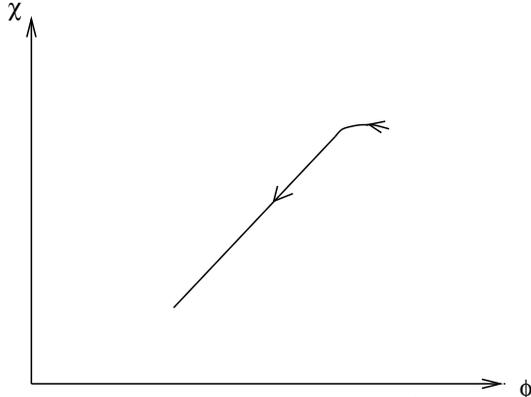,width=7cm}
\caption[h]{Schematic sketch of the evolution of $\phi$ and $\chi$: 
the enhanced friction in brane worlds causes the fields quickly to move 
on a straight line after a short period of damping of the motion of
the angle $\theta$.}
\end{figure}

Note that our findings do not imply that there are no entropy
perturbations generated in brane worlds. In this letter we have 
discussed only the evolution of perturbations {\it during} inflation. 
In principle the power spectra in the radiation dominated epoch 
depend on the details of the decay of the fields $\phi$ and $\chi$ 
and therefore a calculation has to be made using a detailed theory 
of preheating in two--field inflation. 

After inflation, there is another source of entropy perturbation, 
namely the projected Weyl--tensor in equation (\ref{einstein}), 
which acts as a source term for perturbations on 
the brane. It could, in principle, produce entropy perturbations 
even at late times [\ref{langlois}]. However, the amplitude of the 
source term in (\ref{langlois}) for linear perturbations around a 
homogeneous--isotropic brane universe has not been calculated so
far. 

To conclude, we have shown that the modifications to General
Relativity implied by brane world scenarios can suppress the production 
of entropy perturbations during inflation driven by multiple 
scalar fields. The reason is that, for a given potential, 
the background trajectories are less curved in brane worlds due to the 
larger friction acting on the fields. In addition, if inflation is driven at
energies much larger than the brane tension, the final power spectra do
not depend on the initial conditions and/or trajectories of the
fields. This also implies that any transfer of non--Gaussianity 
from the entropy field to the gravitational potential and/or 
adiabatic perturbations is suppressed. Thus, if any entropy
perturbations are detected in the CMB and/or large scale structure, it
would mean that either inflation on a brane happened at energies
smaller than the brane tension or that steeper potentials have 
driven inflation. Although our calculations are based on the
Randall--Sundrum model, we believe that our conclusions carry over 
to other brane world models as well, such as models with a bulk 
scalar field [\ref{lukas}]. In these models, the brane tension is not 
necessarily constant but depends on the evolution of the bulk scalar
field near the brane. As long as the time--evolution of the bulk 
scalar field is negligible compared to the evolution of the inflaton 
fields, our conclusions should hold. Otherwise, there will be a 
source for the curvature perturbation, coming from the apparent 
non--energy--conservation on the brane due to the bulk scalar 
field [\ref{bruck}]. 

\vspace{0.5cm}

{\bf Acknowledgements:} We are grateful for comments from
C. Gordon and G. Rigopoulos. The authors are supported in 
part by PPARC.

\references
\item \label{liddlelyth} A.R. Liddle, D. Lyth, {\it Cosmological
Inflation and Large Scale Structure}, Cambridge University Press
(2000)
\item \label{bellido} J. Garcia-Bellido, D. Wands, 
Phys.Rev. D {\bf 53}, 5437 (1996)
\item \label{durrer} R. Trotta, A. Riazuelo, R. Durrer,
Phys. Rev. Lett. {\bf 87}, 231301 (2001)
\item \label{luca} L. Amendola, C. Gordon, D. Wands, M. Sasaki, 
Phys. Rev. Lett. {\bf 88}, 211302 (2002)
\item \label{gordon} C. Gordon, D. Wands, B.A. Bassett, R. Maartens, 
Phys.Rev.D {\bf 63}, 023506 (2001)
\item \label{nibbelink} S. Groot Nibbelink, B.J.W. van Tent,
Class.Quant.Grav. {\bf 19}, 613 (2002)  
\item \label{bartolo} N. Bartolo, S. Matarrese, A. Riotto, 
Phys.Rev. D{\bf 64}, 123504 (2001)
\item \label{randall} L. Randall, R. Sundrum, Phys.Rev.Lett. {\bf 83}, 
4690 (1999)
\item \label{wands} R. Maartens, D. Wands, B.A. Bassett, I. Heard, Phys.Rev.D 
{\bf 62}, 041301 (2000)
\item \label{liddle} E.J. Copeland, A.R. Liddle, J.E. Lidsey,
Phys.Rev. D {\bf 64}, 023509 (2001)
\item \label{hawkins} R. Hawkins, J. Lidsey, Phys.Rev. D{\bf 63}, 
041301 (2001)
\item \label{lidsey} G. Huey, J. Lidsey, Phys.Lett.B {\bf 514}, 217
(2001); J. Lidsey, T. Matos, L. Urena-Lopez, 
Phys.Rev. D{\bf 66} 023514 (2002)
\item \label{wandsriotto} D. Wands, N. Bartolo, S. Matarrese,
A. Riotto, astro-ph/0205253, to appear in Phys. Rev. D (2002)
\item \label{malik} D. Wands, K. Malik, D. Lyth, A.R. Liddle, 
Phys.Rev. D{\bf 62}, 043527 (2000)
\item \label{shiro} T. Shiromizu, K. Maeda, M. Sasaki, 
Phys.Rev. D{\bf 62}, 024012 (2000)
\item \label{roy} R. Maartens, Phys. Rev.D {\bf 62}, 084023 (2000); 
G. Giudice, E. Kolb, J. Lesgourgues, A. Riotto, hep-ph/0207145 (2002) 
\item \label{langlois} D. Langlois, R. Maartens, M. Sasaki, D. Wands, 
Phys.Rev. D{\bf 63}, 084009 (2001)
\item \label{uzan} F. Bernadeau, J.--P. Uzan, hep-ph/0207295 (2002)
\item \label{lukas} A. Lukas, B.A. Ovrut, K.S. Stelle, D. Waldram,
Phys.Rev. D{\bf 59}, 086001 (1999)
\item \label{bruck} C. van de Bruck, M. Dorca, C.J.A.P. Martins,
M. Parry, Phys. Lett. B{\bf 495}, 183 (2000)
\end{document}